\documentclass[12pt]{article} 
\usepackage[sectionbib]{natbib}
\usepackage{array,epsfig,fancyheadings,rotating}
\usepackage[]{hyperref}  
\usepackage{sectsty, secdot}
\usepackage{subcaption} 
\sectionfont{\fontsize{12}{14pt plus.8pt minus .6pt}\selectfont}
\renewcommand{\theequation}{\thesection\arabic{equation}}
\subsectionfont{\fontsize{12}{14pt plus.8pt minus .6pt}\selectfont}

\usepackage{amsmath}
\usepackage{amssymb}
\usepackage{amsfonts}
\usepackage{multirow}
\usepackage{amsthm}
\usepackage[english]{babel}

\newtheorem{theorem}{Theorem}

\theoremstyle{definition}
\newtheorem{definition}{Definition}

\usepackage[margin=1in]{geometry} 
\begin{document}


\renewcommand{\baselinestretch}{2}

\markright{ \hbox{\footnotesize\rm Statistica Sinica
}\hfill\\[-13pt]
\hbox{\footnotesize\rm
}\hfill }

\markboth{\hfill{\footnotesize\rm LEILA NOMBO AND ANNE-SOPHIE CHAREST} \hfill}
{\hfill {\footnotesize\rm Combining rules for DP datasets} \hfill}

\renewcommand{\thefootnote}{}
$\ $\par


\fontsize{12}{14pt plus.8pt minus .6pt}\selectfont \vspace{0.8pc}
\centerline{\large\bf INFERENCE WITH COMBINING RULES FROM MULTIPLE}
\vspace{2pt} 
\centerline{\large\bf DIFFERENTIALLY PRIVATE SYNTHETIC DATASETS}
\vspace{.4cm} 
\centerline{Leila Nombo, Anne-Sophie Charest} 
\vspace{.4cm} 
\centerline{\it Université Laval}
 \vspace{.55cm} \fontsize{9}{11.5pt plus.8pt minus.6pt}\selectfont


\begin{quotation}
\noindent {\it Abstract:}
Differential privacy (DP) has been accepted as a rigorous criterion for measuring the privacy protection offered by random mechanisms used to obtain statistics or, as we will study here, synthetic datasets from confidential data. Methods to generate such datasets are increasingly numerous, using varied tools including Bayesian models, deep neural networks and copulas. However, little is still known about how to properly perform statistical inference with these differentially private synthetic (DIPS) datasets. The challenge is for the analyses to take into account the variability from the synthetic data generation in addition to the usual sampling variability. A similar challenge also occurs when missing data is imputed before analysis, and statisticians have developed appropriate inference procedures for this case, which we tend extended to the case of synthetic datasets for privacy. In this work, we study the applicability of these procedures, based on combining rules, to the analysis of DIPS datasets. Our empirical experiments show that the proposed combining rules may offer accurate inference in certain contexts, but not in all cases.

\vspace{9pt}
\par
\end{quotation}\par

\def\thefigure{\arabic{figure}}
\def\thetable{\arabic{table}}

\renewcommand{\theequation}{\thesection.\arabic{equation}}

\fontsize{12}{14pt plus .8pt minus .6pt}\selectfont

\section{ Introduction}




The idea of using synthetic datasets for privacy protection has been proposed some thirty years ago (\citet{rubin1993statistical, little1993statistical}), and has since led to a vast literature on methodology to generate and analyze such datasets (see the monograph \cite{drechsler2011synthetic} and a more recent review in \cite{raghunathan2021synthetic}). Software to generate synthetic datasets has also been created (\cite{nowok2016synthpop}, \cite{gaboardi2020programming}), and synthetic datasets have been produced and published by various government organizations (\citet{benedetto2013creation, nowok2017providing}). 

Synthetic datasets have certain advantages over other more traditional statistical disclosure limitation techniques. Notably, once published they allow users to compute any statistics or fit any model of interest, although the accuracy of such outputs of course depends on the relationship between the synthesis model and the model of interest. They also allow more easily for certain tasks which data analysts usually perform on real datasets, such as the testing of modeling assumptions. 

Because they are simulated, synthetic datasets however have the disadvantage of having an additional source of variability, on top of the usual sampling variability. Even if the synthesis model is well chosen, and  the synthetic dataset reassembles the original dataset, it is important to take this source of variability into account as it may have important consequences in the analysis. For example, it is well known in statistics that if variables are measured with error this can lead to attenuation bias in a regression analysis, so that the estimated slope will be biased downward (\cite{Carroll2006}, \cite{Buonaccorsi2010}). Luckily, methods have been designed to handle measurement error. 

Similarly, researchers have proposed methods to obtain valid inference from synthetic datasets, taking into account both the randomness due to the sampling of the original dataset and that introduced by the synthesis model. This task may use sophisticated statistical modeling, but it is also possible to use simple combining rules (\cite{reiter2003inference}) when multiple copies of the synthetic dataset have been published. We'll describe these in more details in section \ref{section_synt_explain}, but the idea here is that we can learn about the variability introduced by the synthesis by comparing the multiple copies of the synthetic datasets generated. These combining rules thus make proper statistical inference accessible to most data users, another important advantage of synthetic datasets. 

In parallel to this work on synthetic datasets, concerns were however raised about the privacy protection offered by these synthetic datasets, and  traditional statistical disclosure control methodology more generally. Several works highlighted possible attacks on statistical products and datasets (\cite{sweeney2000simple}, \cite{dinur2003revealing}). This led to the creation of differential privacy (\cite{dwork2006calibrating}), a formal privacy criterion which is now regarded as the gold standard by many researchers, is used by major technological companies and statistical agencies to handle personal data (\citet{aktay2020google,rogers2020linkedin,applefull}, \cite{censusbureau}). 

Since then, a lot of work has been devoted to the generation of synthetic datasets under the constraint of the formal privacy criterion of differential privacy (\citet{zhang2017privbayes, triastcyn2018generating, raisa2023noise}). Advances have been made in terms of quality of the synthetic datasets, and the complexity of the datasets which can be generated, but little has been done regarding how to properly analyze such datasets. 

There seems to be this idea that if statistics or models from the original dataset and those from the corresponding synthetic datasets are similar enough this is enough. However, as \cite{evans2019statistically} argue here ``a scientific statement is not one that is necessarily correct, but one that comes with known statistical properties and an honest assessment of uncertainty", it is essential to also have a measure of uncertainty.
 
In this paper, we investigate whether the combining rules proposed by statisticians to obtain inference from synthetic datasets can be used in the case of differentially-private synthetic (DIPS) datasets. Early results on this topic showed that the usual combining rules where not appropriate for analysis of synthetic datasets generated with the beta-binomial synthesizer, a simple mechanism to publish counts under differential privacy (\cite{charest2011can}). These results however probably expose the inadequacy of the synthesizer as much as that of the combining rules, and they do not mean that the combining rules could not be  applicable to better synthesizers. 

In fact, \cite{liu2021model} demonstrates that for the bayesian model-based differentially private synthesizer that she designed, the same combining rules can be used to obtain valid inferences, and these were in fact used in subsequent work (\citet{liu2020differentially, chanyaswad2019ron} and \cite{ bowen2020comparative}). While the theoretical analysis of this synthesizer can not be directly extended more generally, the author hypothesized that the same methodology could apply to other synthesizers ``including both model-based and model-free DIPS approaches that do not generate synthetic data from differentially private posterior predictive distributions". 

We empirically test this hypothesis for various differentially private synthesizers, for which the theoretical approach of \cite{liu2021model} is not easily applied. Some of these synthesizers are based on more statistical ideas (bayesian networks, copulas) while others used deep learning models (GANs in particular). Note that the combining rules require generating several synthetic datasets. While this is clearly not ideal from a privacy perspective, as it involves splitting the privacy budget, we argue that this is justified by the fact that we can obtain inferences. If accurate inference could be obtain with a single dataset, we would be happy to do so, but methodology for that has not been published. To limit the privacy cost, we however limit ourselves to five synthetic datasets, whereas some studies my use up to 20. We'll discuss a bit more later this choice. Actually one could generate all five datasets from the same private model. This would also taking into account the variability from the model, but not that of the estimated model itself, which can be quite large under DP. 

We find that the combining rules indeed offer accurate inference in a case where a method produce unbiased or minimal bias for  point estimate and the between variance sufficiently capture variability due do DP. It works for DPGAN method, COPULA-SHIRLEY method and some times for PATE-GAN method.

The paper is organized as follows. Section 2 presents differential privacy and briefly describes the mechanisms to generate DIPS datasets, which will be studied in the paper. Section 3 describes the combining rules used to obtain inference from multiple synthetic datasets which will be tested in the context of differentially private synthesis. Our simulations and results are detailed in section 4, which is followed by a discussion in section 5.  

\section{Differential Privacy}

\subsection{Definition}
Differential privacy (DP) was originally proposed by the seminal paper of \cite{dwork2006calibrating}. Informally, this privacy measure ensures that whether or not a participant agrees to provide their personal information will not have much of an impact on the published output, so that an intruder can not infer personal information about the respondent from the output. Formally, it is defined as follows:
 
\begin{definition} \textbf{($\varepsilon$-Differential Privacy)} Let $\mathcal{D}$ be the set of all possible datasets, and $\varepsilon >0$. A randomized mechanism $\mathcal{K}$ gives $\varepsilon$-differential privacy if, for all datasets $D_1$ and $D_2$ both elements of $\mathcal{D}$ and differing on at most one record, and for all $S \subseteq \mbox{Range} (\mathcal{K})$,  
\begin{equation}
\left| \log \left(\frac{Pr[\mathcal{K}(D_1)\in S] }{Pr[\mathcal{K}(D_2) \in S]  } \right) \right| \leq \varepsilon
\label{def:DP}
\end{equation}

\end{definition}

Equation \ref{def:DP} requires that the distribution of outputs from the mechanism $\mathcal{K}$ does not change too much if one of the observation in the dataset is modified or removed, and that this must hold for any possible dataset. The parameter $\varepsilon > 0$, sometimes called the privacy-loss parameter, controls the desired level of privacy, with smaller values corresponding to greater privacy. Note that this privacy guarantee is thus a property of the randomized algorithm, not of a particular output. 

For any statistical task many different algorithms can be constructed to satisfy DP. In particular, we will present in section \ref{section_gen_dips} several mechanisms whose outputs are completely synthetic datasets. But first, we describe some properties of DP which are important for synthetic data generation, and present a widely-used variant of the original DP definition. For a more complete presentation of DP and examples of mechanisms satisfying DP see \cite{dwork2014algorithmic} and \cite{li2016differential}.


A first important property of DP is its immunity against post-processing.

\begin{theorem} 
\textbf{Post-processing immunity.} Let $\mathcal{K}:\mathcal{D} \to \mathrm{R} $ be a randomized algorithm that satisfies $\varepsilon$-DP. Let $f:\mathrm{R} \to \mathrm{R'}$ be an arbitrary function, independent of $\mathcal{D}$. Then $f \circ \mathcal{K} : \mathcal{D} \to \mathrm{R'} $ also satisfies  $\varepsilon$-DP.
\end{theorem}

This property ensures that it is not possible to reduce the privacy guarantee offered by a mechanism satisfying DP by manipulating a published output. In particular, it means that any statistical analysis can be conducted on DIPS dataset without affecting the privacy guarantee. 

Differential privacy also allows for two different types of composition. 

\begin{theorem}\textbf{Sequential Composition.} Let $\mathcal{K}_1:\mathcal{D} \to \mathcal{R}_1$ and $\mathcal{K}_2:\mathcal{D} \to \mathcal{R}_2$ be two randomized mechanisms that are respectively $\varepsilon_1$ and $\varepsilon_2$ -differentially private. Then the composition $(\mathcal{K}_1,\mathcal{K}_2)(x):\mathcal{D} \to \mathcal{R}_1 \ast \mathcal{R}_2$ is $\varepsilon_1 + \varepsilon_2$ -differentially private.
\end{theorem}

Sequential composition allows to properly quantify the total privacy-loss across multiple uses of an original input dataset. In our context, it is used to conclude that generating $m$ synthetic datasets with a mechanism satisfying $\varepsilon$-DP will have an overall privacy-loss parameter of $m\varepsilon$.

\begin{theorem}\textbf{Parallel Composition}
 Let $\mathcal{K}_1:\mathcal{D}_1 \to \mathcal{R}_1$ and $\mathcal{K}_2:\mathcal{D}_2 \to \mathcal{R}_2$ be two randomized mechanisms that are respectively $\varepsilon_1$ and $\varepsilon_2$ -differentially private such that $\mathcal{D}_1, \mathcal{D}_2 \subset \mathcal{D}$ and $\mathcal{D}_1 \cap \mathcal{D}_2= \emptyset $. Then the composition $(\mathcal{K}_1,\mathcal{K}_2)(x):\mathcal{D} \to \mathcal{R}_1 \ast \mathcal{R}_2$ is $\max(\varepsilon_1, \varepsilon_2)$ -differentially private.
\end{theorem}

Parallel composition thus allows splitting of the input dataset and applying randomized mechanisms to each part while controlling the overall privacy-loss.  

We conclude this section by presenting a relaxation of the original DP definition. This relaxation is widely used to obtain mechanisms with better utility, and in particular is the one satisfied by the deep learning methods we will use to generate synthetic datasets. 
\begin{definition}
 \textbf{($\varepsilon,\delta$- Approximate Differential Privacy)} (\cite{dwork2006our})Let $\mathcal{D}$ be the set of all possible datasets, and $\varepsilon >0$, $\delta >0$. A randomized mechanism $\mathcal{K}$ gives $(\varepsilon,\delta)$-differential privacy if, for all datasets $D_1$ and $D_2$ both elements of $\mathcal{D}$ and differing on at most one element, and for all $S \subseteq \mbox{Range} (\mathcal{K})$, 
  $$Pr[\mathcal{K}(D_1)\in S] \leq exp(\varepsilon) Pr[\mathcal{K}(D_2) \in S]+\delta $$
 \end{definition}
The $(\varepsilon,\delta)$-approximate DP can be interpreted similarly to the pure DP. In this case, it is possible for neighbor datasets to produce some output $S$ with largely different probabilities, but this can only happen for output with small probabilities, as controlled by the value of $\delta$. This parameter must also be chosen by the user, and very small values are usually recommended, such as $10^{-4}$ or $10^{-6}$. 


\subsection{Generation of differentially private synthetic datasets}
\label{section_gen_dips}

There is a very large number of algorithms to generate synthetic datasets while satisfying differential privacy. For this work, we selected six mechanisms for which public code was available and which are representative of a variety of approaches, using both statistical and machine learning models. In the first two cases, code was shared directly by the authors on GitHub. Code for the other four mechanisms was found in the smartnoise-synth Python package of the OpenDP project (\cite{gaboardi2020programming}), so that the implementation may differ a bit from that of the original papers.

Some of algorithms satisfy $\varepsilon$-differential privacy while others satisfy $(\varepsilon,\delta)$-differential privacy.
This is really  not restrictive here since we are interesting on the inference of DIPS datasets not on their generation.

 \textbf{- DataSynthesizer}: 
    We use the algorithm proposed in \cite{ping2017datasynthesizer}, for which code is provided online. The algorithm is based on work from \cite{zhang2014privbayes}. It uses a Bayesian network to approximate the joint distribution of the attributes in the dataset by a product of conditional distributions with lower dimensions and has been shown to satisfy $\varepsilon$-DP. First, an appropriate bayesian network structure is constructed through a modified version of the GreedyBayes algorithm which achieves differential privacy. Next, noisy conditional distributions corresponding to the selected network are obtained, again under the constraint of differential privacy. By default, the overall privacy budget is split equally between these two tasks. Synthetic samples are then generated by sampling from the estimated Bayesian network. We note that in the available code the algorithm automatically discretizes continuous variables and transforms each variable (continuous or categorical) into a set of binary indicators. Improvements have been proposed in \cite{zhang2017privbayes} which offer more advanced techniques to discretize continuous variables and remove the need for binary transformation but we decided to use the first version, for which the provided code is more accessible.

    
\textbf{- COPULA-SHIRLEY}: This method, proposed by \cite{gambs2021growing}, generates DIPS datasets through vine copulas. Copulas (\cite{sklar1959fonctions}) are often used in statistics to model complex multivariate relationship as they allow to estimate the dependence structure between the variables separately from each of their marginal distributions. Vine copulas (\cite{bedford2001probability}, \cite{bedford2002vines}) extend this idea  by using a set of bivariate copulas to estimate the multivariate copula of interest. Such a model can more easily capture complex multivariate relationships. A private vine copulas model is first estimated from the  original dataset and then synthetic data is sampled from it. Half of the dataset is used to obtain noisy histograms from which cumulative density functions are calculated. These private density functions are then used with the other half of the dataset to construct noisy ``pseudo-observations", from which the vine copula model is estimated. Since copulas originally use continuous variables, the algorithm proposes four methods to transform categorical variables, with an inverse processing at the end to obtain synthetic data which are comparable to the original data. The default method, one-hot encoding,is used for our experiments. 
    While \cite{gambs2021growing} shows that the COPULA-SHIRLEY algorithm satisfies $\varepsilon$-DP, we noticed after completing our experiments that there is a mistake in their argument. Discussions with the authors confirmed that in fact the observations used to obtain the ``pseudo-observations" are not protected by differential privacy, but neither of us could find a fix for the problem. We nevertheless included this synthesizer in our simulations as it provides an interesting reference for a mechanism with lesser privacy guarantee.

    
    
    
    
\textbf{- DPGAN}: Generative Adversarial Networks (GANs) (\cite{goodfellow2014generative}) have been recently studied extensively as a powerful tool to generate synthetic datasets (\cite{saxena2021generative}, \cite{chen2020gs}, \cite{torkzadehmahani2019dp}, \cite{xu2019modeling}, \cite{jordon2018pate}, \cite{xu1811synthesizing}). A GAN is a generative model which works by putting two sub-models in competition : a generator which tries to generate new observations similar to those in the original dataset, and a discriminator which tries to identify which of the observations were in the real dataset and which were generated by the generator. Both are trained together until the generator model produces synthetic data hardly distinguishable from the orginal dataset.  \cite{xie2018differentially} proposed DPGAN, a differentially private GAN, obtained by carefully adding noise to the gradients of the discriminator in the learning step. Thus the synthetic datasets generated from the resulting generator model will respect differential privacy through the post-processing property. Note that the DPGAN algorithm achieves ($\varepsilon,\delta$)-approximate DP.

\textbf{- DP-CTGAN}:  CTGAN, for Conditionnal Tabular GAN was proposed by \cite{xu2019modeling} to overcome some challenges when modeling tabular datasets with GANs. More specifically, CTGAN is designed for datasets which contain simultaneously discrete and continuous data, multi-modal and non gaussian continuous variables and severe imbalance in discrete variables. A new normalization method is used to work with non gaussian and multimodal distributions while severe imbalance in discrete variables is handled by using a conditional generator. DP-CTGAN (\cite{fang2022dp}) modifies the  CTGAN method to achieve ($\varepsilon,\delta$)-approximate DP. 

\textbf{- PATE-GAN}: This is another variation on GAN methods, which is based on the PATE (Private Aggregation of Teacher Ensembles) framework (\cite{papernot2016semi}, \cite{papernot2018scalable}), which produces differentially private classifiers. The main idea of PATE-GAN is to use the PATE mechanism as the discriminator in a GAN. To do so, \cite{jordon2018pate} modify slightly the algorithm of PATE to produce a differentiable student model which allows ``back-propagation" to the generator. The PATE-GAN algorithm also achieves ($\varepsilon,\delta$)-approximate DP.    
    
\textbf{- PATE-CTGAN}: This is simply an extension of PATE-GAN using a conditional generator just like DP-CTGAN. While the authors of PATE-GAN (\cite{jordon2018pate}) indicate that the method is able to generate continuous and discrete variables, the function shared by OpenDP does not currently allow the inclusion of continuous variables. Thus, we use PATE-GAN for the discrete data simulations and PATE-CTGAN for the continuous data simulations. 
    


\section{Statistical inference from synthetic data with differential privacy}
\label{section_synt_explain}


This section provides background information on combining rules used to make valid inference from synthetic datasets. The key idea is to generate multiple versions of the synthetic dataset in order to gain information about the added variability from the synthesis process, and to use this information to obtain correct inference. The combining rules are based on methodology created for the analysis of multiply imputed datasets in the case of missing data (\cite{little1993statistical}). They have however been adapted to the privacy case to account for the fact that all observations are imputed, not just a random subset corresponding to the observations with missingness. In fact, the specifics of the combining rules even in the privacy context differ according to how exactly the synthetic data were generated. 


There are two main strategies for creating synthetic datasets, which are clearly described in \cite{drechsler2018some}. In our work, we follow the approach first proposed in \cite{little1993statistical} which consists in estimating a joint model for the observed variables using the observed, private, dataset, and then sampling new observations from this model to generate synthetic data.  In this context, inference should be conducted with the combining rules derived in \cite{reiter2003inference}, which we now describe.


Let $Q$ be the unknown scalar parameter of interest. Using usual statistical methodology, we can obtain from each of the $m$ synthetic datasets a point estimate for $Q$. We can also obtain an estimate of the variance of this estimate, which will help us evaluate the precision of our estimate and can be used to construct a confidence interval for the quantity of interest. Let $q_i$ denote the point  estimate of $Q$ from dataset $i$ and $u_i$ denote its variance estimate, with $i=1 \ldots m$. The overall point estimate for $Q$ is simply the average of the individual point estimates : 
\begin{equation}
\label{eqn:Qhat}
\hat{Q} = \bar{q}_m= \frac{1}{m} \sum_{i=1}^m q_i 
\end{equation}

The variance estimate for this estimator is a little more complex, and involves both the average of the variance estimates from the individual datasets 
\begin{equation}
\bar{u}_m=\frac{1}{m}\sum_{i=1}^m u_i   
\end{equation}
and the variance between the estimates of $Q$ from the different datasets
\begin{equation}
b_m=\frac{1}{m-1} \sum_{i=1}^m (q_i-\bar{q}_m)^2.
\end{equation}
This second term captures the added variability from the synthesis. This added variability has itself two components which are the variability due to DP and the variability due to sampling new synthetic values from the altered model. 
The overall variance estimate of $\hat{Q}$ in equation \ref{eqn:Qhat} is then given by
\begin{equation}
\label{eqn:Tp}
T_p=\frac{b_m}{m} +\bar{u}_m.
\end{equation}
See \cite{reiter2003inference} for the derivation. Note that the $p$ in $T_p$ comes from the fact that the methodology used to generate the synthetic datasets is called \textit{partial synthesis}. This name is a bit unfortunate since we are generating completely synthetic datasets, but it is historically justified. Again, see \cite{drechsler2018some} for details. 
Let $q_{obs}$ denotes the point estimate of $Q$ when using the original dataset and let $u_{obs}$ the usual variance estimate of $q_{obs}$ from the original dataset. For the variance estimator $T_p$ in \eqref{eqn:Tp}, $\bar{u}_m$ estimates $u_{obs}$ and $\frac{b_m}{m}$ estimates the additional variance due to imputations. For inference to be valid with this combining rules estimators, two conditions are required:
\begin{itemize}
    \item the distribution of $q_i$ and $q_{obs}$ are asymptotically normal;
    \item the point estimate $q_i$ is asymptotically unbiased for $q_{obs}$, $b_m$ asymptotically unbiased for $B_0$ where $B_0=E(b_m)$ according to population and $\bar{u}_m$ asymptotically unbiased for $u_{obs}$.
\end{itemize}
Thus, we expect that the combining rules point estimate should be unbiased across values of $\varepsilon$. And for variance estimates, we expect that values increase when $\varepsilon$ decrease due to more noise added. We also hope that the component $b_m/m$ of estimator $T_p$ would properly capture the additional variance due to DP synthesis. 



Note that this variance estimator is exactly the same as the  estimator derived for the differentially private Bayesian synthesis model proposed in \cite{liu2021model} and used in \citet{liu2020differentially, chanyaswad2019ron} and \cite{ bowen2020comparative}. This is thus the estimator that we expect, or hope, will work for the different differentially private synthesis models tested in this paper. 

We will however also consider two other variance estimators : 
\begin{equation}
 T_{s} =\bar{u}_m \left(1+\frac{1}{m}\right)
\end{equation}

\begin{equation}
T_{s(PPD)}=\bar{u}_m \left(1+\frac{2}{m}\right)
\end{equation}

Proposed in \cite{raab2016practical} they are simplifications of $T_p$ in \ref{eqn:Tp}. The estimator $T_s$ should be used when the synthetic data are generated from a estimated model, where $T_{s(PPD)}$ is appropriate when the parameters of these estimates are themselves obtained by sampling from their posterior distribution, so that the synthetic data comes from a posterior predictive distribution (PPD). Both estimators should be  valid for large sample sizes when synthesis is from the same distribution (same family and same parameters) as the one which generated the original dataset. These two estimators even have the advantage of not requiring the generation of multiple synthetic datasets since they involve $\bar{u}_m$ but not $b_m$. This could be useful in the context of confidentiality protection since generating multiple datasets usually requires splitting the privacy budget between the datasets, increasing the noise in each synthetic dataset. However, because of the privacy constraint, the assumptions required for these estimators may not be valid. In fact, simulation results in  \cite{liu2021model} indicate that they were not appropriate for the differentially private synthetic datasets.  We still include them in our experiments to see how they behave with our set of differentially private synthesizers.

We also note that confidence intervals for $Q$ can be easily obtained with all three of the variance estimates. For $T_p$, confidence intervals should be constructed with quantiles from a t-distribution with $v_m$ degrees of freedom where $v_m=(m-1)(1+r_{m_p}^{-1})^2$ and $r_{m_p}=m^{-1}b_m/\bar{u}_m$. For $T_s$ and $T_{s(PPD)}$, a normal distribution is appropriate  when $n$ is large (\cite{drechsler2018some}).

\section{Simulations}

This section presents results for simulations conducted to investigate the validity of the variance estimators discussed in the previous section in the context of differentially private synthesis. Three distinct series of simulations are presented. A first one uses datasets containing only continuous variables with normal distributions. The second one is a variant where one of the variables is highly skewed. The third one uses datasets with only categorical variables.

In all cases, $1000$ original datasets are generated, according to models which will be described below. We set the sample size to $n=10,000$, since some of the more complex synthesizers require larger sample sizes. For each of the $1000$ original datasets, we generate $m=5$ synthetic copies with each of the synthesizers considered, for eight different values of $\varepsilon$ between $0.005$ and $50$. We also include the case of $\varepsilon = 5,000,000$, which represents synthesis with no privacy and is used for comparison purposes. In the case of DPGAN, PATE-GAN and DP-CTGAN, which satisfy approximate differential privacy we use the default value for $\delta$, namely $1/n \sqrt{n}=10^{-6}$. 

Different parameters of interest are considered in the simulations: means, proportions and regression coefficients. Let $\theta$ denote a parameter of interest. We obtain point estimates $\{\hat{\theta}_b \}_{b=1}^{B=1000}$ with the combining rules estimator in equation \ref{eqn:Qhat}. These estimates are compared to their true value in the data generation model, either by computing the bias of the estimates, $\frac{1}{B}\sum_{b=1}^B(\hat{\theta}_b - \theta)$, or the relative bias in percentage, $100 *  \frac{1}{B} \sum_{b=1}^B (\hat{\theta_b}-\theta)/\theta$. 

Variance estimates are also obtained with the three different estimators $T_p$, $T_s$ and $T_{s_{(PPD)}}$, as well as $\bar{u}_m$, which is the naive variance estimate one might compute ignoring the synthesis variability. The true variances of the point estimates are unknown, but we obtain a Monte Carlo estimate from the simulations :
\begin{equation}
V_{MC} = \frac{1}{B-1} \sum_{b=1}^B (\hat{\theta}_b-\bar{\hat{\theta}})^2
\end{equation}
where  $\bar{\hat{\theta}} = \frac{1}{B} \sum_{b=1}^B \hat{\theta}_b$
We then compute what we call a ratio bias, in percentage,
\begin{equation}\label{eqn:rab}
\mbox{RaB} = 100 * \frac{\frac{1}{B} \sum_{b=1}^B \widehat{Var(\hat{\theta})}_b}{V_{MC}}
\end{equation}
where $\widehat{Var(\hat{\theta})}_b$ is an estimate of the variance of the point estimate $\hat{\theta}_b$. In our work, several variance estimators will be used as$\widehat{Var(\hat{\theta})}_b$ : $T_p$, $T_s$, $T_{s_{(PPD)}}$, and $\bar{u}_m$. Values around $100\%$ indicate that the estimator is unbiased, values over $100\%$ that it overestimates the variance and values under $100\%$ that it underestimates the variance. For example, a value of $50$ means that the variance estimator is on average only half of the true variance of the point estimator. We also construct $95\%$ confidence intervals for each parameter using the appropriate distributions and test their coverage.


\subsection{Simulation 1: continuous data with normal variables}

\subsubsection{{\bf Simulation setting}}

For this first simulation, we consider a dataset with three continuous variables $y_1,y_2,y_3$ , generated from a multivariate normal distribution with mean $\mu=(0,0,0)^{'}$, variance $\sigma^2=(1,1,1)^{'}$ and  $Cov(y_1,y_2)=0.8$, $Cov(y_1,y_3)=0.6$ and $Cov(y_2,y_3)=0.25$. As indicated before, a total of $B=1000$ original datasets of size $n=10~000$ are generated. Two parameters are studied in this first simulation: the mean $(E(y_1),E(y_2),E(y_3))^{'}$ whose true value is $(0,0,0)$ and the slopes $\beta_2$ and $\beta_3$ in a multiple linear regression of $Y_1$ on variables $Y_2$ and $Y_3$ defined as: $Y_1=\beta_0 + \beta_2 Y_2 + \beta_3 Y_3 + \gamma, \gamma \sim N(0,1)$. Note that the point estimates for the slopes are compared with a Monte Carlo estimate estimated from the original datasets since their value is not explicit in the data generation process.

The synthesizers used here are DataSynthesizer, COPULA-SHIRLEY, DP-CTGAN, PATE-CTGAN and DPGAN. It is however not possible to use all synthesizers for all values of $\varepsilon$ tested : it was not possible to obtain synthetic datasets with the three smallest values of $\varepsilon$ using the available algorithms for the GANs. On the other end, obtaining a single synthetic dataset with  $\varepsilon=1,000,000$ is so long for the PATE-CTGAN methods that we instead use $\varepsilon=50$ for each dataset, for a total privacy budget of $\varepsilon=250$ instead of $5,000,000$. We should also note again that the COPULA-SHIRLEY algorithm does not quite provide the differential privacy guarantee claimed in the paper. 


\subsubsection{{\bf Results for the means}}

Table \ref{tab:mean_estimates_y1_con} shows the bias of the combining rules estimator $\bar{q}_m$ for the mean of variable $Y_1$ under different scenarios. The estimator is unbiased for DataSynthesizer and COPULA-SHIRLEY for all values of $\varepsilon$.  The bias is negligible for DP-CTGAN for all $\varepsilon$ tested, but for PATE-CTGAN and DPGAN there is some bias for $\varepsilon$ up to  $5$. While we could not obtain results with the GAN methods for the smaller values of $\varepsilon$, they would probably be even more biased. Results for the two others variables are similar and are in the supplementary materials.
\begin{table}[h!]
\centering
\setlength{\tabcolsep}{2pt} 
\renewcommand{\arraystretch}{0.5} 
\begin{tabular}{rrrrrrrrrr}
\hline
  $\varepsilon$& 0.005 & 0.05 & 0.5 & 2.5 & 5 & 8 & 10 & 50 & 5M*\\ 
  \hline
DataSynthesizer & 0.03 & 0.04 & 0.03 & 0.00 & 0.00 & 0.00 & 0.00 & 0.00 & 0.00 \\ 
  COPULA-SHIRLEY & -0.00 & -0.00 & -0.00 & -0.00 & 0.00 & -0.00 & -0.00 & -0.00 & -0.00 \\ 
   DP-CTGAN &  &  &  & 0.14 & 0.00 & 0.01 & 0.02 & -0.04 & -0.03 \\ 
  PATE-CTGAN* &  &  &  & -1.48 & -1.23 & -0.04 & 0.07 & 0.10 & -0.04* \\ 
  DPGAN &  &  &  & 1.88 & 0.56 & -0.07 & -0.02 & -0.06 & -0.14 \\ 
   \hline
\end{tabular}

   *Due to very long running times, these are in fact for $\varepsilon=250$ only for \tiny{PATE-CTGAN}. 
\caption{Bias of the mean estimate of variable $Y_1$ for different values of $\varepsilon$ and different DP synthesis methods for simulation 1.} 
\label{tab:mean_estimates_y1_con}
\end{table}

We now consider the variability of the mean estimates.  Table \ref{sim1_tab:mc_var_con_mean} shows the variance of the mean estimate of variable $Y_1$ over the $1000$ replications for each scenario. Note first that the variance of the estimates is much larger when synthetic data are generated with DPGAN than with any other method, even for values of $\varepsilon$ which gave unbiased estimates of the mean, and in the case of no real privacy ($\varepsilon=1~000~000$). In particular, the variance of the mean estimates is thousands of times larger for DPGAN than for DataSynthesizer for the same privacy protection. For example with $\varepsilon=8$, $95\%$ of values of estimates for the mean are between $-0.04$ and $0.14$ for DataSynthesizer method while there are between $-1.04$ and $2.72$ for DPGAN method.   
For the two more classical synthesizers, the estimates are more variable when $\varepsilon$ decreases, which an be explained by a larger addition of noise to obtain greater privacy. There is however no such trend for the GAN synthesizers. Again, similar results are obtained for $Y_2$ and $Y_3$ and are given in the  supplementary materials.
\begin{table}[ht]
\centering
\setlength{\tabcolsep}{2pt} 
\renewcommand{\arraystretch}{0.5} 
\begin{tabular}{rrrrrrrrrr}
\hline
  $\varepsilon$& 0.005 & 0.05 & 0.5 & 2.5 & 5 & 8 & 10 & 50 & 5M*\\  
  \hline
  DataSynthesizer & 45.23 & 44.68 & 38.23 & 18.76 & 9.68 & 5.41 & 3.96 & 0.34 & 0.11 \\ 
  COPULA-SHIRLEY & 0.22 & 0.26 & 0.25 & 0.24 & 0.21 & 0.20 & 0.18 & 0.15 & 0.15 \\ 
  DP-CTGAN &  &  &  & 547.71 & 6.87 & 5.78 & 7.18 & 40.39 & 72.23 \\ 
  PATE-CTGAN* &  &  &  & 58.41 & 20.74 & 5.64 & 1.80 & 0.37 & 5.69* \\ 
  DPGAN &  &  &  & 2819.63 & 2525.60 & 1806.11 & 2101.04 & 2422.85 & 2407.99 \\ 
   \hline
   \end{tabular}

   *Due to very long running times, these are in fact for $\varepsilon=250$ only for \tiny{PATE-CTGAN}. 
\caption{Variance ($\times 10^3$) of the mean estimates of $Y_1$ for different values of DP parameter $\varepsilon$ and for the synthesis methods for simulation 1.} 
 \label{sim1_tab:mc_var_con_mean}
\end{table}

\begin{table}[h!]
\small
\raggedright
\setlength{\tabcolsep}{2pt} 
\renewcommand{\arraystretch}{0.5} 
\begin{tabular}{rrrrrrrrrr}
  \hline
  $\varepsilon$& 0.005 & 0.05 & 0.5 & 2.5 & 5 & 8 & 10 & 50 & 5M*\\ 
   \hline
DataSynthesizer $T_p$ & 1.82 & 1.81 & 1.81 & 4.04 & 5.41 & 7.23 & 8.31 & 46.53 & 100.31 \\ 
  $T_s$ & 1.31 & 1.32 & 1.45 & 2.25 & 3.48 & 5.20 & 6.51 & 45.78 & 113.70 \\ 
  $T_{s_{ppd}}$& 1.53 & 1.54 & 1.69 & 2.63 & 4.06 & 6.07 & 7.60 & 53.40 & 132.65 \\ 
  $\bar{u}_m$ & 1.09 & 1.10 & 1.21 & 1.88 & 2.90 & 4.33 & 5.43 & 38.15 & 94.75 \\ 
  \hline
  COPULA-SHIRLEY $T_p$ & 102.83 & 99.15 & 101.06 & 94.84 & 95.97 & 95.10 & 96.21 & 98.73 & 95.83 \\ 
  $T_s$ & 53.40 & 45.56 & 47.82 & 49.91 & 57.42 & 61.14 & 65.86 & 82.03 & 81.52 \\ 
  $T_{s_{ppd}}$& 62.30 & 53.15 & 55.79 & 58.23 & 66.99 & 71.33 & 76.83 & 95.70 & 95.10 \\ 
  $\bar{u}_m$ & 44.50 & 37.97 & 39.85 & 41.59 & 47.85 & 50.95 & 54.88 & 68.35 & 67.93 \\ 
  \hline
  DP-CTGAN $T_p$ &  &  &  & 42.00 & 95.00 & 94.47 & 99.01 & 108.04 & 99.45 \\ 
   $T_s$  &  &  &  & 0.15 & 2.12 & 2.89 & 2.49 & 0.57 & 0.23 \\ 
  $T_{s_{ppd}}$  &  &  &  & 0.17 & 2.47 & 3.37 & 2.91 & 0.66 & 0.27 \\ 
  $\bar{u}_m$ &  &  &  & 0.12 & 1.77 & 2.40 & 2.08 & 0.47 & 0.19 \\
  \hline
  PATE-CTGAN* $T_p$  &  &  &  & 325.57 & 553.81 & 84.33 & 129.48 & 59.24 & 88.24* \\ 
   $T_s$  &  &  &  & 1.46 & 4.29 & 10.10 & 22.44 & 32.27 & 2.03* \\ 
  $T_{s_{ppd}}$  &  &  &  & 1.71 & 5.01 & 11.79 & 26.18 & 37.65 & 2.36* \\ 
  $\bar{u}_m$ &  &  &  & 1.22 & 3.58 & 8.42 & 18.70 & 26.89 & 1.69* \\ 
  \hline
  DPGAN $T_p$ &  &  &  & 101.68 & 93.68 & 109.38 & 102.36 & 104.76 & 107.75 \\ 
  $T_s$ &  &  &  & 0.01 & 0.00 & 0.00 & 0.00 & 0.01 & 0.01 \\ 
  $T_{s_{ppd}}$ &  &  &  & 0.01 & 0.00 & 0.00 & 0.00 & 0.01 & 0.01 \\ 
  $\bar{u}_m$ &  &  &  & 0.01 & 0.00 & 0.00 & 0.00 & 0.01 & 0.01 \\ 
   \hline
\end{tabular}

*Due to very long running times, these are in fact for $\varepsilon=250$ only for \tiny{PATE-CTGAN}. 
\caption{Ratio bias, in percentage, for the mean of $Y_1$ for different values of $\varepsilon$ and synthesis methods for simulation 1.} 
\label{sim1_tab:var_estimates_y1_con_moy}
\end{table}

While it is interesting to observe the differences in variances of the mean estimates for different synthesizers, the main question of the paper is whether or not we can estimate this variance correctly with the combining rules. Table \ref{sim1_tab:var_estimates_y1_con_moy} shows the ratio bias measure defined in \ref{eqn:rab}, again for $Y_1$ and all scenarios.  Results for $Y_2$ and $Y_3$ can be found in the supplementary materials with similar conclusions. Among all four estimators tested, $T_p$ clearly is the best one, as expected. It is not however accurate in all cases. It is unbiased for COPULA-SHIRLEY and DPGAN methods for all values of $\varepsilon$ and for DP-CTGAN when $\varepsilon \ge 5$, but it only works for the extreme case of $\varepsilon = 5,000,000$ for DataSynthesizer, and performs inconsistently for PATE-CTGAN. Note that the naive estimator $\bar{u}_m$ consistently underestimates the variance, which shows the importance of considering an alternate estimator to account for the synthesis variability, even when the value of $\varepsilon$ is large.

\begin{table}[h!]
\small
\centering
\setlength{\tabcolsep}{2pt} 
\renewcommand{\arraystretch}{0.5} 
\begin{tabular}{rrrrrrrrrr}
  \hline
  $\varepsilon$& 0.005 & 0.05 & 0.5 & 2.5 & 5 & 8 & 10 & 50 & 5M*\\  
  \hline
  DataSynthesizer $T_p$ & 23.30 & 23.60 & 23.10 & 36.30 & 38.70 & 42.70 & 81.30 & 92.50 & 44.70 \\ 
  $T_s$ & 18.80 & 19.20 & 20.40 & 24.70 & 31.20 & 36.50 & 80.40 & 94.50 & 39.60 \\ 
  $T_{s_{ppd}}$ & 21.10 & 20.60 & 21.80 & 27.00 & 33.80 & 38.50 & 84.00 & 96.50 & 41.80 \\ 
  $\bar{u}_m$ & 18.30 & 18.30 & 19.30 & 24.70 & 30.20 & 35.40 & 77.00 & 91.30 & 37.60 \\ 
     \hline
  COPULA-SHIRLEY $T_p$ & 94.20 & 95.20 & 94.70 & 93.90 & 93.90 & 94.30 & 94.40 & 94.10 & 94.00 \\ 
  $T_s$ & 85.20 & 80.30 & 82.80 & 83.10 & 86.10 & 86.80 & 92.00 & 91.40 & 88.70 \\ 
  $T_{s_{ppd}}$ & 88.10 & 84.00 & 86.30 & 85.80 & 89.50 & 89.10 & 94.00 & 94.90 & 90.90 \\ 
  $\bar{u}_m$ & 85.10 & 80.90 & 83.70 & 83.40 & 85.90 & 85.70 & 89.50 & 89.10 & 87.60 \\ 
     \hline
  DP-CTGAN $T_p$ &  &  &  & 86.70 & 94.80 & 94.30 & 95.50 & 93.90 & 94.40 \\ 
  $T_s$ &  &  &  & 9.20 & 20.50 & 24.60 & 12.20 & 9.40 & 25.70 \\ 
  $T_{s_{ppd}}$ &  &  &  & 9.90 & 22.00 & 26.40 & 13.80 & 10.20 & 27.50 \\ 
  $\bar{u}_m$ &  &  &  & 12.20 & 26.10 & 30.80 & 16.60 & 12.20 & 31.30 \\ 
     \hline
  PATE-CTGAN* $T_p$ &  &  &  & 27.10 & 14.90 & 79.60 & 0.50 & 89.40 & 59.20* \\ 
  $T_s$ &  &  &  & 0.00 & 0.00 & 41.10 & 0.00 & 16.60 & 23.30* \\ 
  $T_{s_{ppd}}$ &  &  &  & 0.00 & 0.00 & 43.90 & 0.00 & 18.60 & 27.20* \\ 
  $\bar{u}_m$ &  &  &  & 0.00 & 0.00 & 41.50 & 0.00 & 21.60 & 29.10* \\ 
     \hline
  DPGAN $T_p$ &  &  &  & 92.20 & 90.40 & 88.10 & 93.90 & 94.90 & 89.50 \\ 
  $T_s$ &  &  &  & 0.80 & 0.40 & 0.50 & 0.90 & 1.00 & 1.00 \\ 
  $T_{s_{ppd}}$ &  &  &  & 0.90 & 0.40 & 0.50 & 0.90 & 1.10 & 1.00 \\ 
  $\bar{u}_m$ &  &  &  & 1.00 & 0.60 & 0.70 & 1.10 & 1.30 & 1.20 \\ 
   \hline
\end{tabular}

   *Due to very long running times, these are in fact for $\varepsilon=250$ only for \tiny{PATE-CTGAN}. 
\caption{Coverage (in \%) of 95 \% confidence intervals for the mean  of $Y_1$, constructed with different variance estimators  for DP synthesis methods for simulation 1.} 
\label{tab:coverage_mean_y1_con}
\end{table}

Table \ref{tab:coverage_mean_y1_con} shows the coverage of 95 \% confidence intervals for the mean  of $Y_1$, constructed with different variance estimators for all scenarios. Results for the two other variables are included in the supplementary materials with similar conclusions. Cases with good coverage generally correspond to those where the variance estimate was accurate : 
COPULA-SHIRLEY for all $\varepsilon$,  DP-CTGAN method when $\varepsilon \ge 5$ and all cases with $\varepsilon = 5,000,000$. For synthetic datasets generated with DPGAN, the coverage is a little lower than expected, which might be due to the small bias seen in table \ref{tab:mean_estimates_y1_con} for the point estimate. For DataSynthesizer method, since all of the variance estimators were biased it is not surprising that they lead to poor coverage.

The derivation of $T_p$ assumes that the component $\bar{u}_m$ is an approximately unbiased estimator for $u_{obs}$, the variance estimator computed on the original dataset. We would thus expect its value not to change with $\varepsilon$. This is tricky because privacy probably changes the distribution, but it may work, at least approximately, for some synthesizers.
\begin{figure}[h!]
\begin{tabular}{cc}
        \includegraphics[angle=0,scale=0.25]{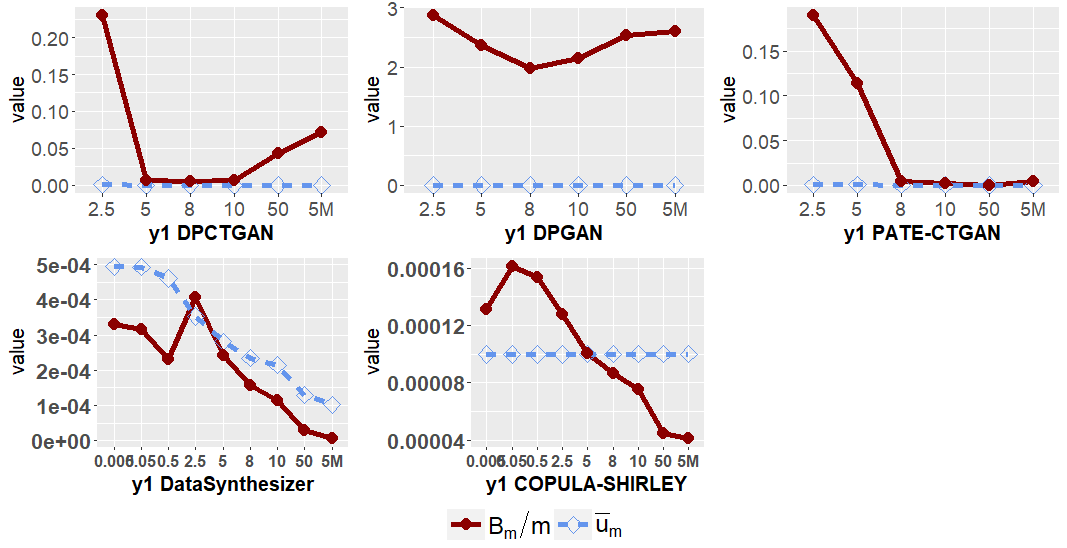}  
&
        \includegraphics[angle=0,scale=0.25]{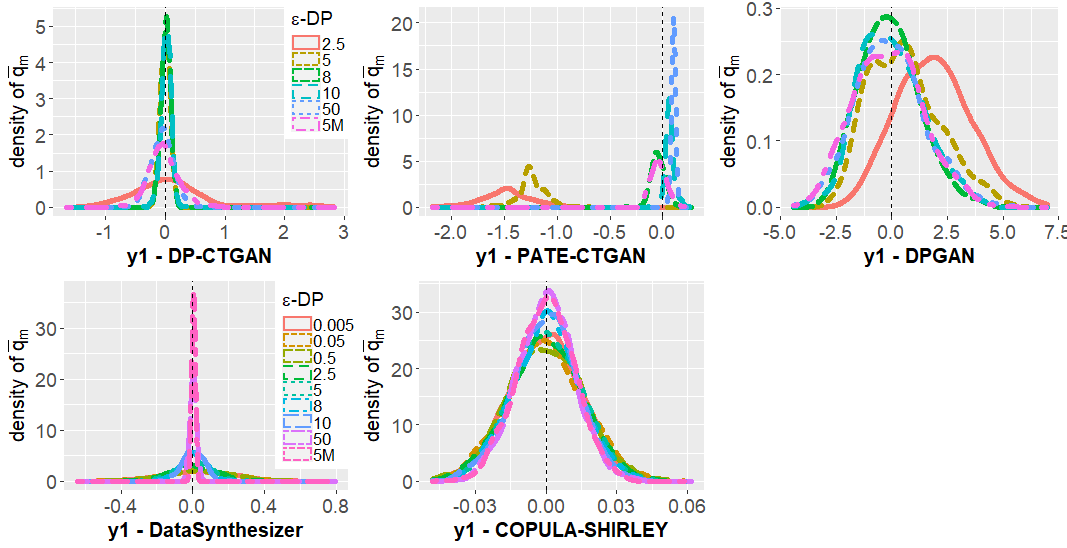}
\end{tabular}    
    \caption{Components of $T_p$ and densities of $\bar{q}_m$ of mean estimate for the variable $Y_1$ over 1000 replications for simulation 1.}
    \label{sim1fig:Bm_var_con_mean}
\end{figure}
We can verify this through figure \ref{sim1fig:Bm_var_con_mean} which presents each components of $T_p$: $\bar{u}_m$ and $B_m/m$ for each value of the DP parameter. We observe that the component $\bar{u}_m$ of $T_p$ does not vary across values of DP parameter $\varepsilon$ for all methods except DataSynthesizer. Only the other component of $T_p$, the between variance $B_m/m$, changes across DP parameter $\varepsilon$. 
The derivation of $T_p$ is also based on each $q_i$ and thus $\bar{q}_m$ following an approximately normal distribution. Figure \ref{sim1fig:Bm_var_con_mean} also shows the densities of $\bar{q}_m$ for the mean of variable $Y_1$ for different values of $\varepsilon$ and synthesis methods. We see that there is an approximately a normal form for densities of $\bar{q}_m$ for COPULA-SHIRLEY, DPGAN and for some values of $\varepsilon$ for DP-CTGAN. This is also true for the two other variables for which densities are in supplementary materials. 


\subsubsection{{\bf Results for slopes for regression}}

We now present results for the estimation of parameters of the following linear regression model: $Y_1=\beta_0+\beta_2 Y_2 + \beta_3 Y_3 + \gamma$ where $\gamma$ represents the model error with a standard normal distribution. Table \ref{sim1_tab:biasslope_estimates_b1_con} shows the relative bias (\%)  of the least square estimates of the slope parameters $\beta_2$ and $\beta_3$. Results for $\beta_0$ are in the supplementary materials. Only COPULA-SHIRLEY produces unbiased estimates for these parameters consistently for all values of $\varepsilon$. The others synthesis methods generally lead to an underestimation of the parameters, with DPGAN being the worst in this case. Even for the case of $\varepsilon=5,000,000$  which corresponds to basically no privacy, only COPULA-SHIRLEY, DataSynthesizer and PATE-CTGAN result in an appropriately small bias.



\begin{table}[ht]
\small
\centering
\setlength{\tabcolsep}{2pt} 
\renewcommand{\arraystretch}{0.5} 
\begin{tabular}{rrrrrrrrrr}
   \hline
  $\varepsilon$& 0.005 & 0.05 & 0.5 & 2.5 & 5 & 8 & 10 & 50 & 5M*\\  
  \hline
$\hat{\beta}_2$  - DataSynthesizer & -101.88 & -101.78 & -99.98 & -91.48 & -81.74 & -72.27 & -67.05 & -27.27 & -2.05 \\ 
  COPULA-SHIRLEY & -0.37 & -0.40 & -0.36 & -0.33 & -0.27 & -0.21 & -0.21 & -0.14 & -0.14 \\ 
  DP-CTGAN &  &  &  & -163.07 & -86.69 & -78.44 & -75.58 & -60.61 & -36.56 \\ 
  PATE-CTGAN* &  &  &  & -95.39 & -74.30 & -85.20 & -58.73 & -4.91 & 5.27* \\ 
  DPGAN &  &  &  & -236.74 & -192.81 & 35.19 & -259.29 & -247.58 & 1001.57 \\ 
  \hline
$\hat{\beta}_3$  - DataSynthesizer & -99.23 & -99.27 & -98.68 & -86.31 & -76.38 & -66.09 & -60.12 & -20.92 & -1.45 \\ 
  COPULA-SHIRLEY & -0.53 & -0.55 & -0.48 & -0.41 & -0.33 & -0.24 & -0.25 & -0.20 & -0.21 \\ 
   DP-CTGAN &  &  &  & -86.20 & -76.93 & -30.96 & -28.06 & -45.77 & -31.87 \\ 
  PATE-CTGAN* &  &  &  & -107.72 & -99.26 & -88.80 & -58.65 & -1.47 & 6.28* \\
  DPGAN &  &  &  & -313.69 & -296.76 & -212.21 & 993.82 & -2024.80 & -462.12 \\ 
   \hline
\end{tabular}

   *Due to very long running times, these are in fact for $\varepsilon=250$ only for \tiny{PATE-CTGAN}. 
\caption{Relative bias (\%) for combining rules point estimates for slopes $\beta_2$ and $\beta_3$ for variables $Y_2$ and $Y_3$ respectively for simulation 1.} 
\label{sim1_tab:biasslope_estimates_b1_con}
\end{table}
Table \ref{sim1_tab:var_estimates_b1_con_reg} shows the ratio bias, in percentage, for the various variance estimators on true variance for $\hat{\beta}_1$. Results for the two other slopes estimates can be found in supplementary materials with similar conclusions. The estimator $T_p$ produces unbiased estimators for all values of $\varepsilon$ for COPULA-SHIRLEY and DPGAN methods. Thus, even in some cases where the point estimators are themselves biased $T_p$ can accurately provide an estimate of the variance of these estimators. $T_p$ also has a small bias for DP-CTGAN for $\varepsilon={2.5, 5}$ and DataSynthesizer for $\varepsilon=5M$. For PATE-CTGAN method where all variance estimates are highly biased except for $\varepsilon={8,50}$. When we look at estimator $\bar{u}_m$, we see again that even for $\varepsilon=5M$, it underestimates variance. This means again that even in the case of no DP, we should apply one correction to variance estimates when we use a synthetic dataset.
\begin{table}[ht]
\centering
\setlength{\tabcolsep}{2pt} 
\renewcommand{\arraystretch}{0.5} 
\small 
\begin{tabular}{rrrrrrrrrr}
  \hline
 $\varepsilon$ & 0.005 & 0.05 & 0.5 & 2.5 & 5 & 8 & 10 & 50 & 5M* \\ 
  \hline
DataSynthesizer $T_p$ & 15244.22 & 14547.19 & 7158.66 & 343.29 & 116.93 & 64.57 & 46.55 & 27.50 & 102.98 \\ 
 $T_s$ & 13560.29 & 12569.35 & 5409.57 & 274.13 & 86.57 & 48.58 & 38.27 & 23.08 & 105.78 \\ 
  $T_{s_{ppd}}$& 15820.34 & 14664.24 & 6311.16 & 319.82 & 101.00 & 56.67 & 44.64 & 26.93 & 123.42 \\ 
  $\bar{u}_m$ & 11300.24 & 10474.46 & 4507.97 & 228.44 & 72.14 & 40.48 & 31.89 & 19.23 & 88.15 \\ 
  \hline
  COPULA-SHIRLEY $T_p$ & 104.92 & 99.13 & 101.14 & 105.39 & 103.70 & 95.42 & 101.80 & 101.55 & 101.65 \\ 
  $T_s$ & 42.06 & 34.89 & 35.77 & 42.00 & 48.65 & 50.77 & 57.60 & 77.46 & 81.18 \\ 
  $T_{s_{ppd}}$& 49.07 & 40.71 & 41.73 & 49.00 & 56.76 & 59.23 & 67.20 & 90.37 & 94.71 \\ 
  $\bar{u}_m$ & 35.05 & 29.08 & 29.81 & 35.00 & 40.54 & 42.31 & 48.00 & 64.55 & 67.65 \\ 
  \hline
  DP-CTGAN $T_p$&  &  &  & 119.72 & 99.56 & 4.11 & 1.26 & 1.70 & 1.27 \\ 
  $T_s$ &  &  &  & 12.95 & 26.56 & 1.01 & 0.29 & 0.11 & 0.06 \\ 
  $T_{s_{ppd}}$ &  &  &  & 15.11 & 30.99 & 1.18 & 0.33 & 0.13 & 0.07 \\ 
  $\bar{u}_m$  &  &  &  & 10.80 & 22.13 & 0.84 & 0.24 & 0.09 & 0.05 \\
  \hline
  PATE-CTGAN* $T_p$ &  &  &  & 162.19 & 422.52 & 97.03 & 171.30 & 101.14 & 42.89* \\ 
  $T_s$ &  &  &  & 56.29 & 24.20 & 14.24 & 60.36 & 116.22 & 19.54* \\ 
  $T_{s_{ppd}}$ &  &  &  & 65.67 & 28.23 & 16.62 & 70.42 & 135.59 & 22.79* \\ 
  $\bar{u}_m$  &  &  &  & 46.91 & 20.17 & 11.87 & 50.30 & 96.85 & 16.28* \\ 
  \hline
  DPGAN $T_p$ &  &  &  & 91.85 & 104.05 & 99.72 & 92.03 & 100.13 & 83.47 \\ 
  $T_s$ &  &  &  & 0.05 & 0.07 & 0.01 & 0.04 & 0.00 & 0.02 \\ 
  $T_{s_{ppd}}$ &  &  &  & 0.06 & 0.08 & 0.02 & 0.05 & 0.00 & 0.02 \\ 
  $\bar{u}_m$ &  &  &  & 0.04 & 0.06 & 0.01 & 0.03 & 0.00 & 0.01 \\ 
   \hline
\end{tabular}

   *Due to very long running times, these are in fact for $\varepsilon=250$ only for \tiny{PATE-CTGAN}. 
\caption{Ratio bias, in percentage, for slope estimate $\hat{\beta}_2$ for different values of $\varepsilon$ and different DP synthesis methods for simulation 1.} 
\label{sim1_tab:var_estimates_b1_con_reg}
\end{table}

For the coverage of the $95\%$ confidence intervals for $\hat{\beta}_2$, since all synthesis methods except COPULA-SHIRLEY produce biased estimators for the slopes, it is not surprising that the nominal coverage of $\hat{\beta}_2$ is achieved only for synthetic data generated with this method. This is also the case for the estimate $\hat{\beta}_3$. Results for variance estimates and coverage are in supplementary materials. 
\begin{figure}[h!]
    \centering
    \includegraphics[angle=0,scale=0.3]{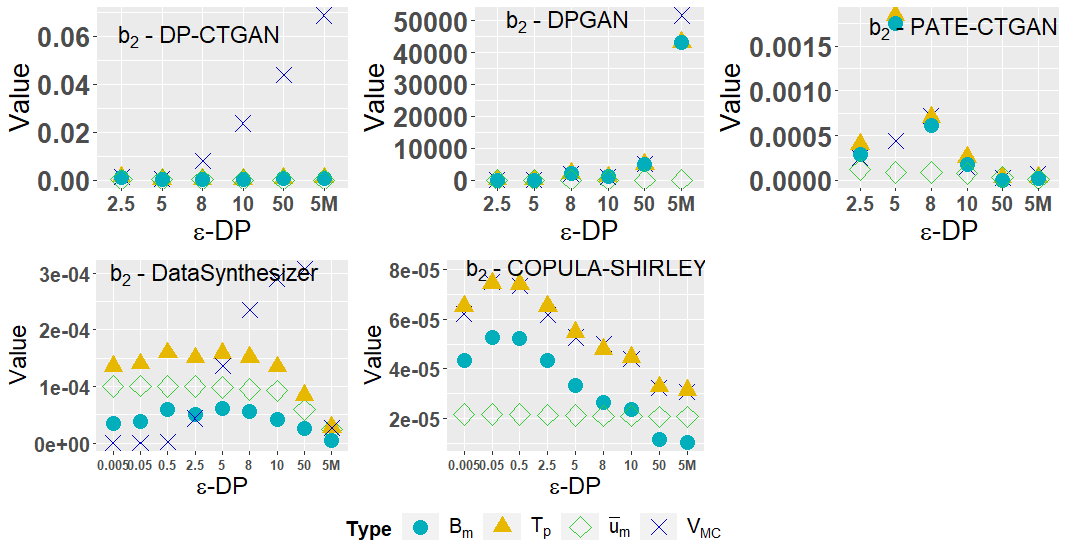}
    \caption{Components of $T_p$, $T_p$ and $V_{mc}$ for estimate of slope for the variable $Y_2$ over 1000 replications for simulation 1.}
    \label{sim1fig:Bm_var_con_reg}
\end{figure}

The derivation of $T_p$ assumes that component $\bar{u}_m$ is approximately unbiased for the variance estimator on the original dataset  $u_{obs}$. Figure \ref{sim1fig:Bm_var_con_reg} shows the trend of each components of the variance estimate $T_p$ for the regression. We see that methods where $T_p$ works are those where $\bar{u}_m$ does not change significantly accross $\varepsilon$ as it is expected. The component $B_m/m$ seems to adequately capture both other sources of variability in the DIPS datasets. Thus, the trend of $B_m/m$ is similar to the trend of variance $V_{MC}$.

\clearpage

\subsection{Simulation 2: simulation with continuous data which contains one variable with a highly skewed distribution.}

\textbf{{\bf Simulation setting}}

We now consider a slight variation on the previous simulation, where one of the variables is not normal, but asymmetric. More precisely, we consider three variables $Y_1,Y_2,Y_3$ with $Y_2\sim N(0,1)$, $Y_3\sim Exp(1)$, $Y_1=1+Y_2+3Y_3+\gamma$ and $\gamma \sim N(0,1)$. To sample from this model, we first simulate two correlated standard normal variables $Y_2$ and $U_3$ with $cov(Y_2,U_3)=0.25$ with the R package \texttt{mvrnorm}. Using the univariate normal cumulative distribution function (CDF) we transform$U_3$ into probabilities, and then apply the inverse CDF of the exponential distribution to obtain values for $Y_3$. 
The rest of the simulation and evaluation is identical to that for the first simulation. To simplify the tables and figures, we however only present results for $T_p$, since the other estimators performed badly again. Complete results are in the supplementary materials.





\textbf{{\bf Results for Mean parameter}}

Table \ref{sim2_tab:summary_y3_con} presents the bias of the mean and the ratio bias for the variance estimate $T_p$ for the normal variable $Y_2$ and the skewed variable $Y_3$. Mean estimates are unbiased or have a slight bias for COPULA-SHIRLEY, DP-CTGAN and  DataSynthesizer, in particular in the case of higher values of $\varepsilon$. However, bias of the mean is a little more marked for the skewed variable $y_3$ than for the normal variable $y_2$ specially when DP parameter $\varepsilon$ decreases. In the case of  DPGAN for example, we see that mean estimate is highly more bias for the skewed variable $y3$ than for the normal variable. As for the variance estimation, $T_p$ performs well for COPULA-SHIRLEY and DPGAN for all values of DP parameter. We observe some loss of precision for the skewed variable $y_3$. This is remarkable for DP-CTGAN where $T_p$ has a slight bias for variance when $\varepsilon \ge 5$ for the normal variable and is biased for the same values of $\varepsilon$ for the skewed variable. The same can be seen for DataSynthesizer when $\varepsilon=5M$.

\begin{table}[h!]
\small
\centering
\setlength{\tabcolsep}{2pt} 
\renewcommand{\arraystretch}{0.5} 
\begin{tabular}{rrrrrrrrrrr}
  \hline
  &$\varepsilon$& 0.005 & 0.05 & 0.5 & 2.5 & 5 & 8 & 10 & 50 & 5M*\\ 
  \hline
   & \multicolumn{10}{c}{\textbf{Bias for mean of normal variable $y_2$ }}\\
   \hline
\multirow{5}{*}{\begin{turn}{90}Bias $y_2$ \end{turn}}  &DataSynthesizer & -0.03 & -0.03 & 0.01 & -0.00 & -0.00 & -0.00 & -0.00 & 0.00 & -0.00 \\ 
  &COPULA-SHIRLEY & -0.00 & -0.00 & -0.00 & -0.00 & -0.00 & -0.00 & -0.00 & -0.00 & -0.00 \\ 
  &DP-CTGAN &  &  &  & 0.46 & 0.05 & 0.03 & 0.03 & -0.00 & -0.00 \\ 
   &PATE-CTGAN* &  &  &  & 1.64 & 0.97 & 0.34 & 0.02 & -0.61 & -1.36* \\ 
  &DPGAN &  &  &  & 3.45 & 0.15 & 0.51 & 0.13 & 0.22 & 0.15 \\ 
   \hline
   & \multicolumn{10}{c}{\textbf{Bias for mean of skewed variable $y_3$}}\\
   \hline
\multirow{5}{*}{\begin{turn}{90}Bias $y_3$\end{turn}}&DataSynthesizer & 3.91 & 3.88 & 3.53 & 2.44 & 1.75 & 1.31 & 1.12 & 0.30 & 0.02 \\ 
 & COPULA-SHIRLEY & -0.00 & -0.00 & -0.00 & -0.01 & -0.01 & -0.01 & -0.01 & -0.01 & -0.01 \\  
 & DP-CTGAN &  &  &  & -0.44 & -0.98 & -0.95 & -0.73 & -0.09 & -0.09 \\ 
  &PATE-CTGAN* &  &  &  & 8.90 & 9.94 & 8.06 & 4.17 & 0.23 & -0.29* \\
 & DPGAN &  &  &  & 10.91 & 7.06 & 2.60 & -0.01 & 2.80 & 2.80 \\ 
   \hline
 & \multicolumn{10}{c}{\textbf{Ratio bias for $T_p$ for mean of normal variable $y_2$}}\\
   \hline
\multirow{5}{*}{\begin{turn}{90}Var. Est. $T_p$ \end{turn}} &DataSynthesizer $T_p$ & 2.15 & 2.06 & 1.71 & 2.50 & 3.65 & 5.10 & 6.29 & 41.07 & 108.53 \\ 
  &COPULA-SHIRLEY $T_p$ & 99.57 & 96.64 & 102.96 & 98.95 & 96.82 & 98.15 & 100.70 & 99.57 & 105.73 \\ 
   &DP-CTGAN $T_p$ &  &  &  & 18.45 & 98.18 & 105.12 & 97.12 & 98.26 & 102.47 \\ 
  &PATE-CTGAN* $T_p$ &  &  &  & 384.93 & 349.39 & 21.50 & 87.30 & 5.24 & 1.61* \\ 
 &DPGAN $T_p$ &  &  &  & 102.07 & 93.26 & 93.59 & 102.52 & 95.89 & 96.80 \\ 
   \hline
   
& \multicolumn{10}{c}{\textbf{Ratio bias for $T_p$ for mean of skewed variable $y_3$}}\\
   \hline
\multirow{5}{*}{\begin{turn}{90}Var. Est. $T_p$ \end{turn}} &DataSynthesizer $T_p$ & 0.24 & 0.25 & 0.33 & 0.67 & 1.15 & 1.68 & 2.03 & 9.89 & 77.74 \\ 
&COPULA-SHIRLEY $T_p$ & 106.08 & 97.46 & 95.64 & 94.17 & 101.75 & 99.80 & 92.20 & 96.06 & 98.93 \\ 
&DP-CTGAN $T_p$ &  &  &  & 3.21 & 3.21 & 78.09 & 9.03 & 71.54 & 79.97 \\ 
&PATE-CTGAN* $T_p$  &  &  &  & 195.98 & 122.11 & 169.05 & 112.77 & 5.28 & 3.18* \\ 
&DPGAN $T_p$ &  &  &  & 97.38 & 97.25 & 100.39 & 98.88 & 92.28 & 93.59 \\ 
   \hline
\end{tabular}
\footnotesize{*According to too long time running, the last column is for $\varepsilon=250$ only for \tiny{PATE-CTGAN}.}
\caption{Bias of the mean and Ratio bias for the variance estimate $T_p$ for variables $Y_2$ and $Y_3$ for simulation 2 with a highly skewed variable. } 
\label{sim2_tab:summary_y3_con}
\end{table}

\begin{figure}[h!]
\begin{tabular}{cc}
        \includegraphics[angle=0,scale=0.25]{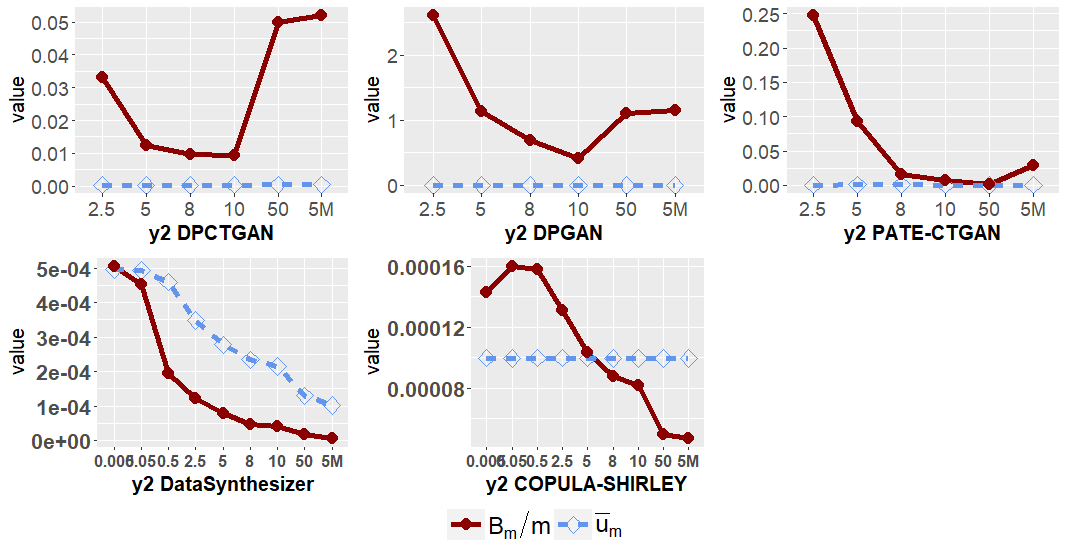}     
&
        \includegraphics[angle=0,scale=0.25]{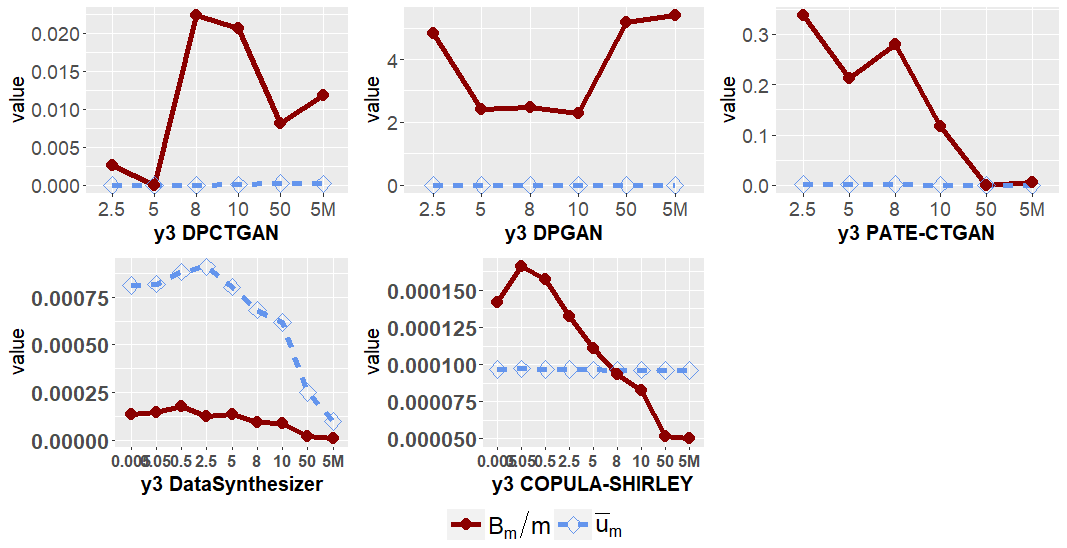} 
\end{tabular}
\caption{Mean of components of $T_p$: $B_m/m$ and $\bar{u}_m$ for mean for normal variable $Y_2$ and skewed variable $Y_3$ over 1000 replications for simulation 2.}
  \label{sim2_fig:bmvm_meany2y3}   
\end{figure}
To investigate more on conditions of use of estimator $T_p$, we present in figure \ref{sim2_fig:bmvm_meany2y3} the mean of components of estimator $T_p$: $\bar{u}_m$ and $B_m/m$ over $1~000$ replications. We observe that component $\bar{u}_m$ does not vary so much across values of DP parameter for all DP synthesis methods except for DataSynthesizer. In inverse, component $B_m/m$ increases when parameter $\varepsilon$ diminishes for COPULA-SHIRLEY, PATE-CTGAN and DataSynethsizer. This is not surprising since component $B_m/m$ is the one which should capture the other types of variability (variability due to DP, due to synthesis process). However, we can see normal form for densities of mean estimate for the normal variable $y_2$ for all values of DP parameter for COPULA-SHIRLEY and for some values of DP-parameter for GANs methods. For the skewed variable, only COPULA-SHIRLEY still produce normal form for densities of $\bar{q}_m$. We can found distribution of $\bar{q}_m$ for mean for those variables in supplementary materials.



\textbf{{\bf Results for slopes for regression}}

Table \ref{sim2_tab:biasslope_estimates_con} shows the relative bias (\%) for estimates and ratio bias (\%) for variance estimates for slopes of regression of $Y_1$ on $Y_2$ and $Y_3$. $\hat{\beta}_2$ and  $\hat{\beta}_3$ are estimates for slopes of normal variable and skewed variable respectively. Except for COPULA-SHIRLEY method, slopes estimates are biased for both variables for all values of $\varepsilon$. Bias has a decreasing linear trend only for DataSynthesizer method. For variance estimator $T_p$, it still works for COPULA-SHIRLEY and DPGAN with a minimal bias. We notice again a small loss of precision for the skewed variable compared to the normal variable mostly for COPULA-SHIRLEY and DP-CTGAN. 
\begin{table}[h!]
\centering
\setlength{\tabcolsep}{2pt} 
\renewcommand{\arraystretch}{0.5} 
\small
\begin{tabular}{rrrrrrrrrr}
  \hline
 $\varepsilon$& 0.005 & 0.05 & 0.5 & 2.5 & 5 & 8 & 10 & 50 & 5M* \\
   \hline
 \multicolumn{10}{c}{\textbf{Relative bias (\%) for $\hat{\beta}_2$ of normal variable $y_2$}}\\
 \hline
 DataSynthesizer & -106.32 & -107.23 & -101.44 & -92.24 & -80.85 & -67.70 & -64.06 & -14.73 & 1.96 \\ 
  COPULA-SHIRLEY & -11.56 & -10.77 & -11.33 & -12.76 & -13.56 & -14.34 & -14.63 & -15.89 & -15.70 \\ 
  DP-CTGAN &  &  &  & -104.19 & -95.23 & -98.79 & -95.28 & -74.46 & -65.26 \\ 
  PATE-CTGAN* &  &  &  & -101.85 & -96.74 & -98.53 & -83.90 & -29.99 & -41.60* \\
  DPGAN &  &  &  & -450.06 & -1796.15 & -1285.68 & 4532.58 & 2527.39 & 4825.48 \\ 
   \hline
   \multicolumn{10}{c}{\textbf{Relative bias (\%) for $\hat{\beta}_3$ of skewed variable $y_3$}}\\
 \hline
   DataSynthesizer & -99.42 & -98.35 & -85.13 & -61.54 & -52.34 & -47.11 & -44.73 & -28.29 & -2.20 \\ 
  COPULA-SHIRLEY & -1.00 & -1.06 & -0.93 & -0.85 & -0.75 & -0.76 & -0.77 & -0.75 & -0.61 \\ 
  DP-CTGAN &  &  &  & -68.61 & 5675.97 & 62789.87 & 18215.41 & -58.88 & -43.62 \\ 
  PATE-CTGAN* &  &  &  & -103.48 & -107.23 & -111.02 & -93.51 & -24.95 & 3.17* \\ 
  DPGAN &  &  &  & 83.20 & -91.17 & -136.37 & -105.61 & -137.29 & -131.83 \\ 
   \hline
    
       \multicolumn{10}{c}{\textbf{Ratio bias (\%) of $T_p( \hat{\beta}_2 )$ of normal variable $y_2$}}\\
          \hline
DataSynthesizer & 4832.23 & 3168.87 & 358.36 & 53.70 & 42.07 & 44.87 & 43.19 & 75.00 & 87.78 \\ 
  COPULA-SHIRLEY & 103.08 & 95.21 & 103.45 & 96.38 & 99.66 & 92.37 & 101.49 & 95.94 & 95.83 \\ 
  DP-CTGAN &  &  &  & 56.44 & 105.79 & 95.27 & 69.76 & 4.55 & 6.05 \\ 
  PATE-CTGAN* &  &  &  & 629.27 & 956.34 & 2243.42 & 605.82 & 8.47 & 11.60* \\ 
  DPGAN &  &  &  & 97.02 & 98.98 & 99.01 & 93.39 & 100.95 & 99.92 \\ 
   \hline
   \multicolumn{10}{c}{\textbf{Ratio bias (\%) of $T_p( \hat{\beta}_3 )$ of skewed variable $y_3$}}\\
      \hline
DataSynthesizer & 48548.27 & 4238.46 & 136.61 & 23.05 & 19.84 & 20.57 & 21.31 & 22.52 & 48.96 \\ 
  COPULA-SHIRLEY & 93.98 & 91.58 & 98.63 & 98.44 & 92.94 & 92.57 & 82.86 & 82.96 & 83.79 \\ 
  DP-CTGAN &  &  &  & 31.62 & 103.82 & 89.01 & 27.75 & 36.61 & 32.66 \\ 
  PATE-CTGAN* &  &  &  & 473.30 & 4147.45 & 277.11 & 217.13 & 18.51 & 66.27* \\ 
  DPGAN &  &  &  & 109.87 & 105.52 & 103.39 & 94.20 & 84.66 & 66.93 \\ 
   \hline
\end{tabular}

   *Due to very long running times, these are in fact for $\varepsilon=250$ only for \tiny{PATE-CTGAN}. 
\caption{Relative bias (\%) and Ratio bias (\%) of $T_p$ for the slopes estimate $\hat{\beta}_2$ and $\hat{\beta}_3$ for different values of $\varepsilon$ and different DP synthesis methods for simulation 2 of continuous data with a skewed variable.} 
\label{sim2_tab:biasslope_estimates_con}
\end{table}

Similarly to simulation 1, we observe that component  $\bar{u}_m$ does not vary very much across values of $\varepsilon$ for all methods except for DataSynthesizer; and the cases where $T_p$ does not work, are those where component $B_m/m$ fails to properly capture the additionnal variance due to DP synthesis process. Figure for components of $T_p$ for this regression can be found in supplementary materials.

\subsection{Simulation 3: binary data}

\subsubsection{{\bf Simulation setting}}
This third simulation is a dataset of three binary variables, again with size $n=10~000$ and where $p(y_1=1)=p(y_2=1)=p(y_3=1)=0.6$; $Cov(y_1,y_2)=Cov(y_1,y_3)=0.6$ and $Cov(y_2,y_3)=0.2$. For that, we use function \texttt{rmvbin} from \texttt{bindata} package where $p(y_1=1)=p(y_2=1)=p(y_3=1)=0.6$ and with the following binary correlations $\rho(y_1,y_2)=\rho(y_1,y_3)=0.6$ and $\rho(y_2,y_3)=0.2$. We are interested in estimating the probabilities of success $p(y_1=1)=p(y_2=1)=p(y_3=1)$ using the combining rule estimate based on the sample proportion as well as estimating the accuracy of the variance estimates. 

\subsubsection{{\bf Results for the probability of success}}

Table \ref{tab:bias_prop_bin} shows the relative bias (in \%) of the combining rule estimate based on the sample proportion for variable $Y_1$ for different values of differential privacy parameter $\varepsilon$ and for different synthesis methods. Results for the other variables are in the supplementary materials with similar conclusions.  The point estimate $\bar{q}_m$ is unbiased for COPULA-SHIRLEY and DataSynthesizer for all values of $\varepsilon$-DP except a slight bias at  $\varepsilon=0.005$. Among GANs methods, DPGAN is the best with a small bias lower than $6\%$. However, PATE-GAN generally produces biased estimates and underestimates the point estimate for all values of $\varepsilon$-DP, this is also true for $Y_2$ and $Y_3$ as can be seen in the supplementary materials. Relative bias in case of DP-CTGAN method is  quite variable : the size and direction of the bias depending on both the variable of interest and the value of $\varepsilon$. Bias presents in point estimate for PATE-GAN method and in some cases for DP-CTGAN method should be corrected to produce good inference.
\begin{table}[h!]
\centering
\setlength{\tabcolsep}{2pt} 
\renewcommand{\arraystretch}{0.5}
\begin{tabular}{rrrrrrrrrr}
  \hline
  $\varepsilon$ & 0.005 & 0.05 & 0.5 & 2.5 & 5 & 8 & 10 & 50 & 5M* \\ 
  \hline
  DataSynthesizer & -8.64 & 2.45 & 0.17 & 0.26 & 0.27 & 0.27 & 0.27 & 0.27 & 0.25 \\ 
  COPULA-SHIRLEY & -3.73 & 0.15 & 0.04 & 0.02 & 0.02 & 0.02 & 0.03 & 0.03 & 0.05 \\ 
  DP-CTGAN &  &  &  & 29.56 & -1.23 & 7.04 & -13.50 & 0.11 & 0.02 \\
  PATE-GAN* &  &  &  & -16.86 & -16.81 & -16.88 & -16.68 & -17.18 & -16.64* \\ 
  DPGAN &  &  &  & -3.99 & -5.49 & -3.81 & -5.46 & -0.85 & -1.19 \\ 
   \hline
\end{tabular}

   *Due to very long running times, these are in fact for $\varepsilon=250$ only for \tiny{PATE-GAN}. 

\caption{Relative bias (in \%) of of the combining rule estimate based on the sample proportion for variable $Y_1$ for simulation 3.}
\label{tab:bias_prop_bin}
\end{table}

\begin{table}[h!]
\small
\centering
\setlength{\tabcolsep}{2pt} 
\renewcommand{\arraystretch}{0.5}
\begin{tabular}{rrrrrrrrrr}
  \hline
 $\varepsilon$& 0.005 & 0.05 & 0.5 & 2.5 & 5 & 8 & 10 & 50 & 5M* \\ 
  \hline
DataSynthesizer $T_p$ & 54346.66 & 657.96 & 153.23 & 131.69 & 130.61 & 130.54 & 130.34 & 130.01 & 117.75 \\ 
  $T_s$& 257.04 & 125.59 & 121.40 & 122.93 & 123.44 & 123.87 & 123.85 & 124.02 & 119.36 \\ 
  $T_{s_{ppd}}$& 299.88 & 146.52 & 141.64 & 143.42 & 144.02 & 144.51 & 144.49 & 144.69 & 139.26 \\ 
  $\bar{u}_m$ & 214.20 & 104.66 & 101.17 & 102.44 & 102.87 & 103.22 & 103.21 & 103.35 & 99.47 \\ 
  \hline
  COPULA-SHIRLEY $T_p$ & 101.11 & 94.20 & 99.30 & 101.17 & 99.40 & 99.49 & 100.74 & 95.42 & 101.22 \\ 
  $T_s$ & 0.09 & 7.15 & 77.86 & 86.33 & 85.35 & 85.20 & 86.67 & 82.07 & 86.84 \\ 
  $T_{s_{ppd}}$ & 0.11 & 8.34 & 90.84 & 100.72 & 99.57 & 99.40 & 101.12 & 95.74 & 101.31 \\ 
  $\bar{u}_m$ & 0.08 & 5.96 & 64.89 & 71.94 & 71.12 & 71.00 & 72.23 & 68.39 & 72.37 \\
  \hline
  DP-CTGAN $T_p$ &  &  &  & 2.79 & 8.04 & 6.48 & 47.35 & 103.36 & 95.20 \\ 
  $T_s$ &  &  &  & 0.27 & 0.07 & 0.07 & 0.29 & 2.96 & 2.73 \\ 
  $T_{s_{ppd}}$ &  &  &  & 0.32 & 0.08 & 0.08 & 0.34 & 3.46 & 3.19 \\ 
  $\bar{u}_m$ &  &  &  & 0.23 & 0.06 & 0.06 & 0.24 & 2.47 & 2.28 \\ 
  \hline
  PATE-GAN* $T_p$ &  &  &  & 107.70 & 98.58 & 98.00 & 102.58 & 98.29 & 103.05* \\ 
  $T_s$ &  &  &  & 0.44 & 0.41 & 0.43 & 0.44 & 0.12 & 0.01* \\ 
  $T_{s_{ppd}}$ &  &  &  & 0.51 & 0.48 & 0.50 & 0.51 & 0.15 & 0.01* \\ 
  $\bar{u}_m$ &  &  &  & 0.36 & 0.34 & 0.36 & 0.36 & 0.10 & 0.01* \\ 
  \hline
  DPGAN $T_p$ &  &  &  & 95.25 & 104.49 & 101.90 & 101.42 & 96.27 & 95.04 \\ 
  $T_s$ &  &  &  & 0.02 & 0.01 & 0.02 & 0.02 & 0.04 & 0.04 \\ 
  $T_{s_{ppd}}$ &  &  &  & 0.02 & 0.02 & 0.02 & 0.03 & 0.04 & 0.04 \\ 
  $\bar{u}_m$ &  &  &  & 0.02 & 0.01 & 0.02 & 0.02 & 0.03 & 0.03 \\ 
   \hline
\end{tabular}

   *Due to very long running times, these are in fact for $\varepsilon=250$ only for \tiny{PATE-GAN}. 
\caption{Ratio bias, in percentage, for of the combining rule estimate based on the sample proportion $Y_1$ for different values of $\varepsilon$ and different DP synthesis methods for simulation 3.} 
\label{tab:varbiasbin}
\end{table}

Table \ref{tab:varbiasbin} shows the ratio bias (in \%) for the variance estimates of the of the combining rule estimate based on the sample proportion for different values of $\varepsilon$ and  synthesis methods. As for the previous simulations, the estimator $T_p$ is the best one. It works well for all the synthesis methods, except for DP-CTGAN and  DataSynthesizer. Surprisingly, it is the variance estimate $\bar{u}_m$ which is unbiased in the case of DataSynthesizer method, at long as$\varepsilon \ge 0.05$; and the estimator $T_p$ overestimate the variance in all cases for this synthesizer. 

\begin{table}[h!]
\small
\centering
\setlength{\tabcolsep}{2pt} 
\renewcommand{\arraystretch}{0.5}
\begin{tabular}{rrrrrrrrrr}
  \hline
 $\varepsilon$& 0.005 & 0.05 & 0.5 & 2.5 & 5 & 8 & 10 & 50 & 5M* \\ 
  \hline
   DataSynthesizer $T_p$ & 100.00 & 100.00 & 98.00 & 96.70 & 96.40 & 96.40 & 96.50 & 95.60 & 96.40 \\  
  COPULA-SHIRLEY $T_p$ & 94.99 & 95.70 & 94.20 & 94.30 & 95.30 & 95.20 & 95.30 & 95.70 & 95.40 \\  
  DP-CTGAN $T_p$ &  &  &  & 2.10 & 36.70 & 66.60 & 96.00 & 95.50 & 27.30 \\ 
  PATE-GAN* $T_p$ &  &  &  & 85.00 & 84.50 & 83.20 & 89.60 & 92.80 & 83.60* \\  
  DPGAN $T_p$ &  &  &  & 91.40 & 91.97 & 91.51 & 92.36 & 91.76 & 91.94 \\ 
    \hline
\end{tabular}

*Due to very long running times, these are in fact for $\varepsilon=250$ only for \tiny{PATE-GAN}. 

\caption{Coverage (in \%) of 95 \% confidence intervals for $T_p$ for the of the combining rule estimate based on sample proportion of $Y_1$ for simulation 3.} 
\label{tab:covproy1bin}
\end{table}

We investigate now the coverage (in \%) of 95 \% confidence intervals of variance estimate $T_p$ for of the combining rule estimate based on the sample proportion. Table \ref{tab:covproy1bin} shows results for variable $y_1$ and results for the other variables are included in the supplementary materials with similar conclusions. Overall, the intervals based on $T_p$ perform relatively well for COPULA-SHIRLEY and DPGAN for all values of $\varepsilon$ and for DP-CTGAN when $\varepsilon={50, 5M}$. Values of the coverage probability for those methods are around $95\%$ except for DPGAN when is around $92\%$ due surely to the minimal bias observed for the point estimate $\bar{q}_m$ in table \ref{tab:bias_prop_bin}. Intervals for DataSynthesizer suffer from a bit of overcoverage for $T_p$.  For PATE-GAN method, values for 95\% coverage probability are far below and up to 84\% when $\varepsilon \le 10$. This is mostly due to the bias of $\bar{q}_m$ observed in table \ref{tab:bias_prop_bin}.
\begin{figure}[h!]
   \begin{tabular}{cc}
         \includegraphics[angle=0,scale=0.23]{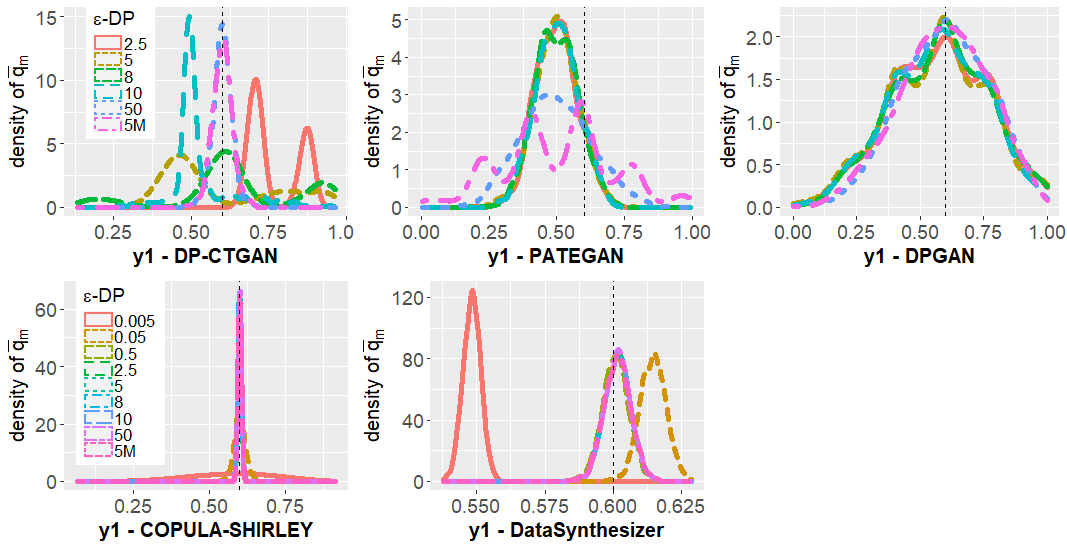}     
&
        \includegraphics[angle=0,scale=0.23]{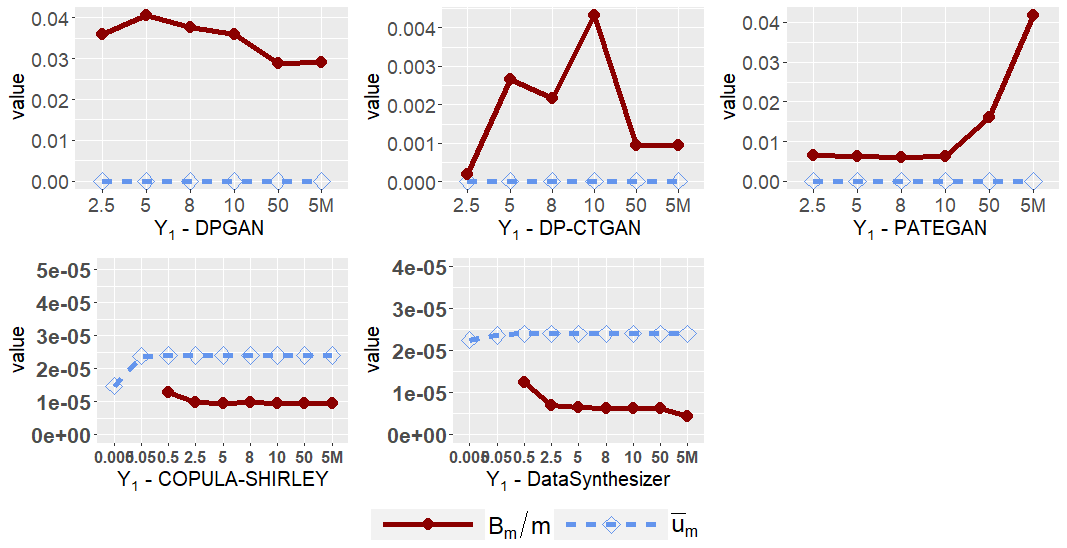}
\end{tabular}
 \caption{Densities for estimator $\bar{q}_m$ and mean of components of variance estimate $T_p$: $B_m/m$ and $\bar{u}_m$ for the combining rule estimate based on the sample proportion  of variable $Y_1$ for simulation 3.}
 \label{sim2_fig:bmmvm_propy1}
\end{figure}
We want now to argue a little bit on conditions for the estimator $T_p$ which works well for some methods (COPULA-SHIRLEY, DPGAN, PATE-GAN) and not for other (DataSynthesizer). Figure \ref{sim2_fig:bmmvm_propy1} shows the densities for $\bar{q}_m$ and the mean of components $\bar{u}_m$ and $B_m/m$ for variance estimator $T_p$ of the combining rule estimate based on the sample proportion  of variable $Y_1$, over 1000 replications and across different values of DP parameter $\varepsilon$ and synthesis methods. We can see globally an approximate normal form for each density and for each synthesis method. However, there are some curves which are not well centered on the true value of the probability of success ($p=0.6$). As observed on previous simulations and as we expect, components $\bar{u}_m$ does not vary much accross values of DP parameter $\varepsilon$ also here. The component $B_m/m$ is the one which vary to capture the other types of variability.

\clearpage
\section{Discussion}

Our original research question was whether or not combining rules proposed for inference with synthetic datasets for confidentiality protection could be used also with synthesizers designed to achieve differential privacy. While previous results concluded that they were not accurate for the beta-binomial synthesizer \cite{charest2011can}, more recent tests with a Bayesian model-based differential private synthesizer showed that combining rules from \cite{reiter2003inference} also applied in that case \cite{liu2021model}. Our goal was to verify if they could be applied in general to DP synthesizers, including more complex ones using deep learning. 

The answer to this question turns out to be complicated. The variance estimator $T_p$ applied very well to synthesis with COPULA-SHIRLEY, leading to low bias for the variance of the point estimate and appropriate coverage in all our experiments. However, as indicated before, this synthesizer does not quite offer the DP guarantee intended. The results were also mostly positive with DPGAN and PATE-GAN methods. Increasing the number of synthetic datasets could improve the estimates, but it would also require adding more noise to each DIPS dataset. This tradeoff should be investigated in future research. For certain synthesis methods however, such as DataSynthesizer and DP-CTGAN, none of the variance estimators could really provide accurate inference. While we offered some discussion as to why this might be the case, more work is needed to fully understand under which conditions $T_p$ is applicable. 

While this was not the main focus of our work, our simulations also highlighted that not all synthesizers produce synthetic data from which we can obtain accurate estimates of means and regression parameters, even in the simple settings considered here. For the continuous dataset with normal variables, only COPULA-SHIRLEY led to unbiased estimates of the regression parameters, and some of the GAN synthesizers could not produce unbiased estimates of the proportions of interest in the binary variables dataset. This failure of the GAN synthesizers here should not be taken as proof that they are inferior for synthetic data generation however since we used the default parameters provided in each case. More tuning might have been necessary to obtain better generators, but this tuning process is trickier under the constraint of differential privacy, and little advice on how to do so is offered. 

Note however that the accuracy of $T_p$ as a variance estimate does not depend on the accuracy of the synthesis model used to generate the data. The variance estimation process really is about quantifying the uncertainty in the estimate, not the quality of it. For example, consider the estimate of the parameters $\hat{\beta}_2$ and  $\hat{\beta}_3$ for the dataset with continuous variables. Table \ref{sim1_tab:biasslope_estimates_b1_con} shows that DPGAN produces highly biased estimates but $T_p$ can still relatively accurately measure the variance of the estimates, which we can see in table \ref{sim1_tab:var_estimates_b1_con_reg}. 

Our simulations also highlight the fact that the variability added during the synthesis is very different for the different synthesizers. COPULA-SHIRLEY provides much less variable estimates in all scenarios, along with PATE-CTGAN for our continuous datasets and Datasynthesizer for our binary datasets. On the other hand, DPGAN leads to more variable estimators for all cases tested. Of course less variability is preferred, whether or not that variability can be accurately quantified with $T_p$ and so these differences should be further investigated. Certain models might be preferred with more complicated datasets than the ones used in our simulations. Simulation including a highly skewed variable also shows that there is a small loss of variability when using that type of variables. 

The use of simple datasets is definitely a limitation of our study in general, and future work should investigate these question on real datasets closer to those used in practice. Still, our work offers a first analysis of the applicability of inference with combining rules from multiple synthetic datasets generated with differential privacy, and shows that it might be a good solution in some cases.

\section*{Supplementary Material}
The supplementary material includes the results for all others variables and estimators of parameter which have not been presented here for all simulations.
\par
\section*{Acknowledgements}

The work of both authors was partially supported by the Natural Sciences and Engineering Research Council of Canada grant No. RGPIN-435472-2013, FONCER 528124-2019, and a CANSSI Collaborative Research Team Project. 
\par


\bibhang=1.7pc
\bibsep=2pt
\fontsize{9}{14pt plus.8pt minus .6pt}\selectfont
\renewcommand\bibname{\large \bf References}
\expandafter\ifx\csname
natexlab\endcsname\relax\def\natexlab#1{#1}\fi
\expandafter\ifx\csname url\endcsname\relax
  \def\url#1{\texttt{#1}}\fi
\expandafter\ifx\csname urlprefix\endcsname\relax\def\urlprefix{URL}\fi

 \bibliographystyle{chicago}      
  \bibliography{Bibliographie.bib}   


\vskip .65cm
\noindent
Université Laval
\vskip 2pt
\noindent
E-mail: (leila.nombo.1@ulaval.ca)
\vskip 2pt

\noindent
Université Laval
\vskip 2pt
\noindent
E-mail: (anne-sophie.charest@mat.ulaval.ca)

\end{document}